\documentclass[prd,12pt,onecolumn,tightenlines,superscriptaddress,numerical,notitlepage,showpacs,amsmath,amssymb,amsfonts,aps,longbibliography,floatfix
]{revtex4-1}
\usepackage{graphicx}
\usepackage{subcaption}
\usepackage[colorlinks=True,linkcolor=red,citecolor=blue,urlcolor=blue]{hyperref}
\usepackage{bm}
\usepackage{bbm}
\usepackage{array}
\usepackage{braket}
\usepackage{cancel}
\usepackage{xcolor}
\usepackage{slashed}
\usepackage{comment}
\usepackage{combelow}
\usepackage{chngcntr}
\usepackage{nicefrac}
\usepackage{bookmark}
\usepackage[T1, T2A]{fontenc}
\usepackage[english]{babel}
\usepackage[normalem]{ulem}

\newcommand{\Tr}{\mathop{\mathrm{Tr}}}

\begin{document}
\selectlanguage{english} 

\title{Cassini Ansatz in the Skyrme Model}

%
\author{
\firstname{N.~V.}~\surname{Gerasimeniuk}}
\email[E-mail: ]{gerasimenyuk.nv@dvfu.ru}
\affiliation{Pacific Quantum Center, Far Eastern Federal University, Vladivostok 690922, Russia}
%
\author{\firstname{A.~A.}~\surname{Korneev}}
\affiliation{Beijing Institute of Mathematical Sciences and Applications, Huairou District, Beijing 101408, China}
\affiliation{Yau Mathematical Sciences Center, Tsinghua University, Beijing 100084, China}
%
\author{\firstname{A.~V.}~\surname{Molochkov}}
\affiliation{Pacific Quantum Center, Far Eastern Federal University, Vladivostok 690922, Russia}
\affiliation{Beijing Institute of Mathematical Sciences and Applications, Huairou District, Beijing 101408, China}
%
\author{\\\firstname{O.~G.}~\surname{Tkachev}}
    \thanks{Deceased.}
\affiliation{Pacific Quantum Center, Far Eastern Federal University, Vladivostok 690922, Russia}
%

\begin{abstract}
\noindent \textbf{Abstract} --- We investigate the stretching of a spherically symmetric solution in the Skyrme model. The three-dimensional generalization of the Cassini ovals is considered as ansatz of extension. The linear dependence of static energy on the longitudinal deformation parameter is obtained with a slope $\sigma \approx 0.14\ \mathrm{ GeV}^2$, which is in good agreement with the string tension known from various models of confinement.
\end{abstract}

\maketitle

\vspace{-5mm}

\section{Introduction}
The Skyrme model, proposed by Tony Skyrme in the 1960s, is an effective theory of strong
interactions based on chiral symmetry and topological solitons — skyrmions. In this model,
baryons emerge as stable configurations of pion fields with non-trivial topology. One of the
key aspects of the model is its ability to describe confinement by topological constraints -- the phenomenon of quark
confinement within hadrons, which in QCD is associated with the formation of a linear potential
at large distances. Initially, the Skyrme model was investigated in the context of a spherically symmetric configuration of pion fields. The study demonstrated how an object with a baryon number, finite size, and a nontrivial topology can arise from a simple assumption of field theory. Our paper presents a revision of the Skyrme model based on a three-dimensional extension of Cassini ovals. This assumption allows us to study how a spherical skyrmion deforms, thereby broadening our understanding of hadron structure in the Skyrme model.

\section{Skyrme model}
The $SU(2)$ Skyrme model~\cite{Skyrme_1961, Skyrme_1962, Adkins_1983, Nikolaev_1990} is defined by the Lagrangian density (\ref{lagrangian}),
written in Euclidean space (using the system of units $\hbar=c=1$):
\begin{eqnarray}
{\cal L}=\frac{f_\pi^2}{4}\Tr\left(\mathrm{L}_\mu \mathrm{L}_\mu\right)+\frac{1}{32e^2}\Tr\left[\mathrm{L}_\mu, \mathrm{L}_\nu\right]^2,
\label{lagrangian}
\end{eqnarray}
where $\mathrm{L}_\mu =\partial_\mu UU^+$ — left-invariant Cartan forms, expressed via $SU(2)$ matrices $U~=~U({\bf x},t)=e^{i{\boldsymbol{\tau}}\cdot{\boldsymbol{\pi}}({\bf x},t)/f_\pi}$, 
defined by an isotriplet of pseudoscalar fields $\boldsymbol{\pi}({\bf x},t)$, which are associated
with $\pi$~--~meson fields, 
$\boldsymbol{\tau}=(\tau_1,\tau_2,\tau_3)$ is the vector of Pauli matrices. The $f_\pi$ is the pion decay constant and $e$ is dimensionless parameter, which was introduced by Skyrme to stabilize solution~\cite{Skyrme_1962}.

The Lagrangian (\ref{lagrangian}) defines a chiral-invariant theory. The Lagrangian is invariant under the global $SU(2)_L\times SU(2)_R$ chiral transformation group $U({\bf x})\to A U({\bf x}) B^+$. This symmetry implies conserved isotopic axial and vector currents. The model has a conserved non-Noether current — a topological current that is identified with the baryon current \cite{Adkins_1983}:
\begin{eqnarray}
{\cal{J}}^B_\alpha=-\frac{1}{24\pi^2}\varepsilon_{\alpha\mu\nu\rho}\Tr\left(\mathrm{L}_\mu \mathrm{L}_\nu \mathrm{L}_\rho\right).
\label{Qa}
\end{eqnarray}
The integral over the volume of the zero component of the current $J^B_\alpha$ is the baryon charge:
\begin{eqnarray}
B=\int\limits_V d{\bf x}\ {\cal{J}}^B_0=-\frac{1}{24\pi^2}\varepsilon_{ijk}\int\limits_V d{\bf x}\ \Tr\left(\mathrm{L}_i \mathrm{L}_j \mathrm{L}_k\right).
\label{B}
\end{eqnarray}

We use the most general assumption about the $SU(2)$ parameterization of the $U$ matrices:
\begin{eqnarray}
U({\bf x}, t)=\exp\Big\{i{\boldsymbol{\tau}}\cdot{\boldsymbol{\pi}}({\bf x},t)/f_\pi\Big\}=\cos F({\bf x},t)+i\left({\boldsymbol{\tau}}\cdot{\bf{N}}({\bf x},t)\right) \sin F({\bf x},t),
\label{U}
\end{eqnarray}
expressed in terms of the modulus
$F({\bf x},t)=\sqrt{{\boldsymbol{\pi}}({\bf x},t)\cdot{\boldsymbol{\pi}}({\bf x},t)}/f_\pi=\vert{\boldsymbol{\pi}}({\bf x},t)\vert/f_\pi$ of the triplet of pseudoscalar fields
${\boldsymbol{\pi}}({\bf x},t)$ and the unit pseudovector ${\bf{N}}({\bf x},t)={\boldsymbol{\pi}}({\bf x},t)/\vert{\boldsymbol{\pi}}({\bf x},t)\vert$, 
defined in the isotopic space. Lagrangian density (\ref{lagrangian}) and the baryon current density (\ref{Qa}) in terms of the functions $F=F({\bf x},t)$ and ${\bf{N}}={\bf{N}}({\bf x},t)$ \cite{Nikolaev_1990}:
\begin{eqnarray}
{\cal L} = &-\dfrac{f_\pi^2}{2}&
\bigg\{\partial_\mu F\partial_\mu F+\left(\partial_\mu{\bf{N}}\cdot\partial_\mu{\bf{N}}\right)\sin^2F\bigg\}
\label{lagrFN}\\
&-\dfrac{1}{2e^2}&
\left\{\left[\partial_\mu{\bf{N}}\times\partial_\nu{\bf{N}}\right]^2\sin^2F+
2\left(\partial_\mu{\bf{N}}\cdot\partial_\nu{\bf{N}}\right)\partial_\alpha F\partial_\beta F
\Delta_{\mu\nu\alpha\beta}
\right\}\sin^2F,
\nonumber
\end{eqnarray}
\begin{eqnarray}
{\cal{J}}^B_\alpha=-\frac{1}{4\pi^2}\varepsilon_{\alpha\mu\nu\rho}
\left(\left[\partial_\mu{\bf{N}}\times\partial_\nu{\bf{N}}\right]\cdot{\bf{N}}\right)
\partial_\rho F\sin^2F,
\label{QaFN}
\end{eqnarray}
where $\delta_{\alpha\beta}$ is the Kronecker delta and $\Delta_{\mu\nu\alpha\beta}=\delta_{\mu\nu}\delta_{\alpha\beta}-\delta_{\mu\alpha}\delta_{\nu\beta}$.

\section{Coordinate system}

\begin{figure}[!ht]
\begin{tabular}{c c c}
  \includegraphics[width=.2\linewidth]{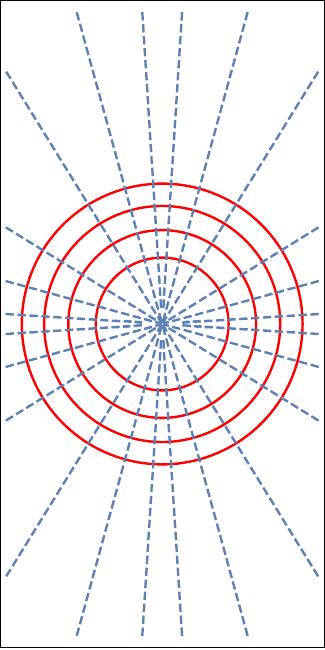} & 
  \includegraphics[width=.2\linewidth]{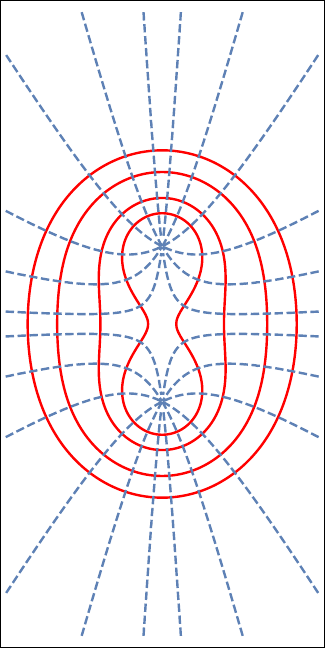} &
  \includegraphics[width=.2\linewidth]{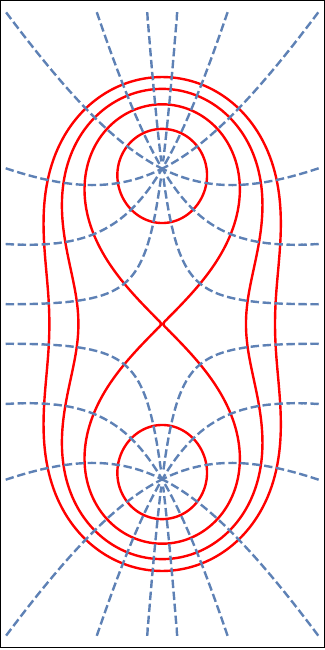} 
\end{tabular}
\caption{Contours of Cassini ovals (solid) and orthogonal lines to them (dashed) for several values of $a=\{0.0,\ 0.5,\ 1.0\}$.}
\label{fig_CasNorm}
\end{figure}

Let us attempt to consider the splitting of a spherically symmetric skyrmion into two parts along the $x_3$ axis of the Cartesian coordinate system. Our ansatz is based on the assumption that this skyrmion transforms into a pair of deformed skyrmions upon splitting. As a coordinate system we use three-dimensional generalization of Cassini ovals (also known as Cassinian curve~\cite{bronshtein_2015_handbook}).
Thus, we assume that the isotopic vector ${\bf{N}}$ in the parametrization~(\ref{U}) of the field $U({\bf x})$ is normal to the Cassini surface $P({\bf x})=\textrm{Const}$ (Fig. \ref{fig_CasNorm}) and the modulus of the pion field $F({\bf x})=F(P)$.

Based on the ideas discussed above, we have chosen a new coordinate system
\begin{eqnarray}
P^2   = 
\sqrt{x_1^2+x_2^2+(x_3-a)^2}
\sqrt{x_1^2+x_2^2+(x_3+a)^2},\; 
\Phi  =  \mathrm{atan}{\left(\frac{x_2}{x_1}\right)},\;
\mathcal{Z}  =  x_1^2+x_2^2+x_3^2,
\label{CgCS}
\end{eqnarray}
where $P>0$, $0\le\Phi<2 \pi$, ${\cal Z}>0$ and $a$ is half the distance between the foci $\{0,0,\pm a\}$ of the Cassini surfaces. Here we have a one-to-one correspondence between the set $\left\{x_1,x_2,x_3\right\}$ and $\left\{P,\Phi,{\cal Z}\right\}$ for $a \neq 0$. The inverse transformations divide configuration into two regions: the upper ($x_3>0$) and the lower ($x_3<0$):
\begin{eqnarray}
x_1=\frac{\sqrt{P^4-\left(a^2-{\cal Z}\right)^2}}{2a}\cos{\Phi},
x_2=\frac{\sqrt{P^4-\left(a^2-{\cal Z}\right)^2}}{2a}\sin{\Phi},
x_3=\pm\frac{\sqrt{\left(a^2+{\cal Z}\right)^2-P^4}}{2 a}
\label{xToP}.
\end{eqnarray}
The Jacobians of the transition from the Cartesian coordinate system to the new coordinate system (\ref{CgCS}) and the change in the variable ${\cal Z}$ over the regions have the form:
\begin{eqnarray}
J_{\pm}=\pm\frac{P^3}{2a\sqrt{\left(a^2\pm\mathcal{Z}\right)^2-P^4}},
\ \mathcal{Z}\in\left[\vert a^2\mp P^2\vert,\vert a^2\pm P^2\vert\right]\ \textrm{for}\ x_3\gtrless 0.
\label{Jac}
\end{eqnarray}

\section{Results}
In the new coordinate system (\ref{CgCS}) we can obtain the following expressions for the construction which are included in the Lagrangian density and density of baryon charge:
\begin{eqnarray}
\partial_\mu P\partial_\mu P&=&\frac{{\cal Z}}{P^2},
\label{dPdP}\\
\left(\partial_\mu{\bf{N}}\cdot\partial_\mu{\bf{N}}\right)&=&\frac{\left(3{\cal Z}^2-P^4+2{\cal Z}a^2+3 a^4\right)}{{\cal Z}P^4},
\label{dNdN}\\
\left[\partial_\mu{\bf{N}}\times\partial_\nu{\bf{N}}\right]^2&=&\frac{2\left(a^2+{\cal Z}\right)^2\left(2a^4+2{\cal Z}^2-P^4\right)}{{\cal Z}^2P^8},
\label{dNxdN}\\
\left(\partial_\mu{\bf{N}}\cdot\partial_\nu{\bf{N}}\right)\partial_\alpha P\partial_\beta P
\Delta_{\mu\nu\alpha\beta}&=&
\frac{a^8\hspace{-1mm}+8a^2{\cal Z}^3\hspace{-1mm}+13{\cal Z}^4\hspace{-1mm}-6{\cal Z}^2P^4\hspace{-1mm}+P^8\hspace{-1mm}+2a^4\hspace{-1mm}\left(5{\cal Z}^2\hspace{-1mm}-P^4\right)}{4{\cal Z}^2P^6},
\label{dFdNdNdF}
\end{eqnarray}
where we imply $\mu,\nu={1,2,3}$ in the static case.

The density of baryon charge for Cassini ansatz can be obtained from expression (\ref{QaFN}), in variables (\ref{CgCS}) it looks as follows:
\begin{eqnarray}
{\cal{J}}^B_0=-\frac{1}{12\pi}\frac{3\left({\cal Z}+a^2\right)\left(3{\cal Z}^2-P^4+a^4\right)}{{\cal Z}^{3/2} P^5}\sin^2F(P)F'(P).
\label{J0}
\end{eqnarray}
By integrating ${\cal{J}}^B_0$ over ${\cal Z}$ and $\Phi$, we obtain the density of baryon charge in the form:
\begin{eqnarray}
Q(P)=\int\limits_0^{2\pi}d\Phi\left(\int\limits_{\vert a^2-P^2\vert}^{a^2+P^2}d{\cal Z}~J_{+}{\cal{J}}^B_0+
\int\limits_{a^2+P^2}^{\vert a^2-P^2\vert}d{\cal Z}~J_{-}{\cal{J}}^B_0\right)=
\left\{ 
\begin{array}{ll}
Q_\textrm{int}^{+}(P)+Q_\textrm{int}^{-}(P), & P<a \\ \\
Q_\textrm{out}(P), & P\ge a
\end{array}
\right. 
\label{J0B}
\end{eqnarray}
and
\begin{eqnarray}
Q_\textrm{out}(P)=Q_\textrm{int}^{+}(P)=-Q_\textrm{int}^{-}(P)=-\frac{2}{\pi}\sin^2F(P)F'(P).
\label{QB}
\end{eqnarray}

\begin{figure}[!ht]
\includegraphics[width=0.7\linewidth] {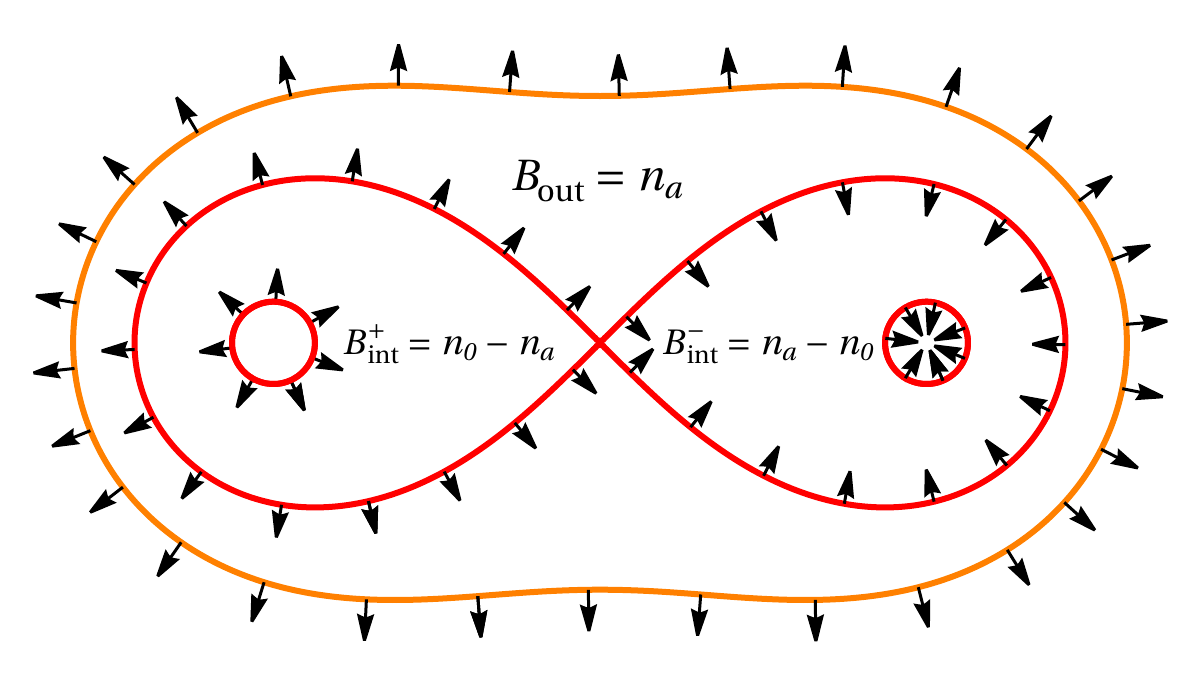}
  \caption{Schematic distribution of baryon charge across different regions and vectors orthogonal to surfaces $P=\textrm{Const}$.}
  \label{fig_Bn}
\end{figure}

The expression $Q_\textrm{int}^{+}(P)=-Q_\textrm{int}^{-}(P)$ reflects the fact that in the inner region ($P<a$) for $x_3>0$ we have a right-oriented coordinate system, and for $x_3<0$ -- a left-oriented one.
The integral over $P$ for expression (\ref{QB}) is taken analytically regardless of the explicit form of the function $F(P)$.
The result depends only on the values $F(0)=n_0\pi$, $F(a)=n_a\pi$ and $F(P)\vert_{P\to\infty}=0$. 
By integrating $Q(P)$, we arrive at the following formula for baryon charge:
\begin{eqnarray}
B=B_\textrm{int}^{+}+B_\textrm{int}^{-}+B_\textrm{out}=B_\textrm{out}=n_a.
\label{Bn}
\end{eqnarray}
Thus, in the inner region ($0\le P<a$) there is a baryon charge $B_\textrm{int}=0$, which is divided into two parts: in the upper ($x_3>0$) inner region the baryon charge is $B_\textrm{int}^{+}=(n_0-n_a)$ -- this is a skyrmion, and in the lower ($x_3<0$) inner region the baryon charge is $B_\textrm{int}^{-}=-(n_0-n_a)$ -- this is an anti-skyrmion.
In the outer region ($P\ge a$) the baryon charge is $B_\textrm{out}=n_a$. This configuration is schematically shown in Fig.~\ref{fig_Bn}.

\begin{figure}[!ht]
\begin{tabular}{c c}
  \includegraphics[width=.3\linewidth]{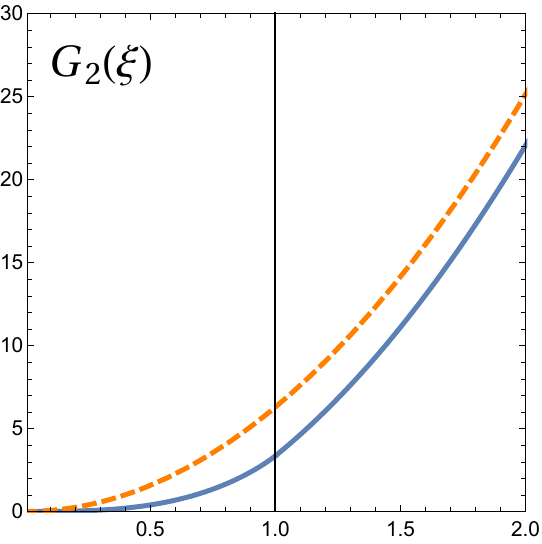} & $\;$
  \includegraphics[width=.3\linewidth]{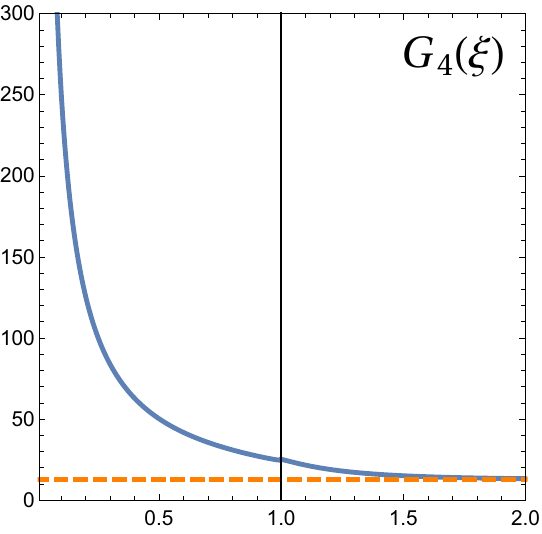} \\
  \includegraphics[width=.3\linewidth]{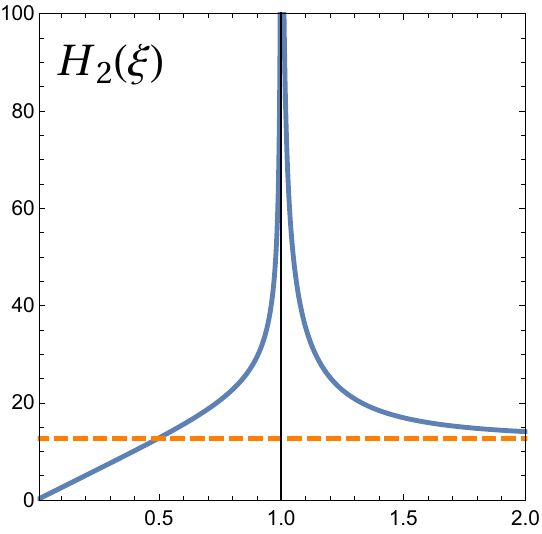} & $\;$
  \includegraphics[width=.3\linewidth]{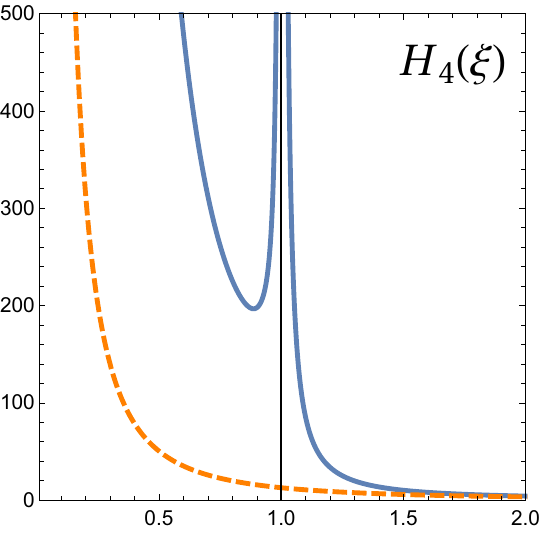}
\end{tabular}
  \caption{Behavior of individual Lagrangian terms in dimensionless variables $\xi=e f_\pi P$. The solid curve shows the terms for which the Cassini ansatz was used. The dashed line indicates similar terms obtained by T. Skyrme for a spherically symmetric ansatz \cite{Skyrme_1962}. The vertical central line indicates $\eta=e f_\pi a = 1$.}
  \label{fig_G2_G4_H2_H4}
\end{figure}

The Lagrangian density (\ref{lagrFN}) can also be rewritten using coordinate transformations (\ref{xToP}). 
Taking into account (\ref{Jac}) and (\ref{dPdP})-(\ref{dFdNdNdF}), we obtain the Lagrangian in dimensionless variables $\xi=e f_\pi P$ and $\eta=e f_\pi a$:
\begin{equation}
L=
-\frac{f_\pi}{e}\int\limits_{0}^{\infty}d\xi\left[G_2(\xi)F'(\xi)^2+ \Big(G_4(\xi)F'(\xi)^2
+H_2(\xi)
+\frac{1}{2}H_4(\xi)\sin^2F(\xi)\Big)\sin^2F(\xi)\right].
\label{lagrF}
\end{equation}
Expressions $G_2$, $H_2$, $G_4$, and $H_4$ are not difficult to obtain analytically, but due to their large size, we present them only graphically in Fig.~\ref{fig_G2_G4_H2_H4}. It is worth noting that in the limit $\eta\to 0$ all obtained results will corresponds to the spherically symmetric skyrmion~\cite{Adkins_1983}. As can be seen from Fig.~\ref{fig_G2_G4_H2_H4}
we have singularities at the points $\xi=0$, $\xi=\eta$ and $\xi\to\infty$. They impose restrictions on the class of the final solutions. We must require the following conditions on the function $F(\xi)$ to be satisfied:
\begin{eqnarray}
\left\{F(0)=\pi n_0,\ F(\eta)=\pi n_\eta,\ F(\xi)\vert_{\xi\to\infty}=\pi n_\infty\right\} \quad \text{for} \quad n_0, n_\eta, n_\infty \in \mathbb{Z}.
\label{BCF}
\end{eqnarray}
The expression for Lagrangian (\ref{lagrF}) indicates that any solution $F(\xi)$ can be shifted by $n\pi$. Therefore, without loss of generality, we will assume that $n_\infty=0$. 

In the end, we have the Euler--Lagrange equation for the static case:
\begin{eqnarray}
0 & = & 
\left[G_2(\xi)+G_4(\xi)\sin^2{F(\xi)}\right]F''(\xi)+\left[G_2'(\xi)+G_4'(\xi)\sin^2{F(\xi)}\right]F'(\xi) 
\nonumber\\
& - &\left[H_2(\xi)+H_4(\xi)\sin^2{F(\xi)}-G_4(\xi)F'(\xi)^2\right]\sin{F(\xi)}\cos{F(\xi)}.
\label{EulerF}
\end{eqnarray}

As shown in~\cite{Manton_1987}, the spherically symmetric solution provides the minimum of static mass only for the sector with baryon charge equal to one. In this paper, we consider the deformation of a spherically symmetric skyrmion and find solutions with $B=1$, i.e. $n=n_0=n_{\eta}=1$.

We look for solutions with the minimum static mass $M=-L$. Obviously, in the inner region ($0\le\xi<\eta$) we should obtain a trivial solution $F(\xi)=\pi$. Integration over $\xi$ in the formula (\ref{lagrF}) should be performed on the interval $\left[\eta,\infty\right]$. So, in the inner region only meson objects are produced $B_\textrm{int}^{+}=-B_\textrm{int}^{-}=0$.

Equation (\ref{EulerF}) was solved numerically using the fourth-order Runge-Kutta method. For different interfocal distances $2a$~(\ref{CgCS}), one can calculate the static mass $M$. Then we choose the relative distance between the centers of inertia $z_c$ of the two halves of our object as the deformation parameter $l$:
\begin{eqnarray}
    l = 2 (z_c - z_0),\ \ \textrm{and}\ \ z_c = \frac{\int z \, \mathcal{L}({\bf x}) \, dV}{\int \mathcal{L}({\bf x}) \, dV},
    \label{eq:center_of_inertia}
\end{eqnarray}
with $z_0 = z_c(a)\Big|_{a \to 0}$. As a result, we obtain the dependence of static mass $M$ on longitudinal stretching $l = 2 (z_c - z_0)$. In the case of $l > 1$ this dependence is almost linear and very similar to the behavior of a string. 

If one converts the dimensionless distance $l$ into units of $\mathrm{GeV}^{-1}$: $l\to f_\pi e l$, and the mass $M$ into units of $\mathrm{GeV}$: $M\to M f_\pi/ e$, then we obtain the tension coefficient of the resulting effective string $\sigma\simeq 34.38 f_\pi^2$.
It should be noted that in the static case of the chiral-invariant Skyrme model, the tension coefficient $\sigma$ does not depend on the geometric parameter $e$ included in the second term of Skyrme Lagrangian (\ref {lagrangian}) and depends only on the fundamental constant $f_\pi$. 

\begin{figure}[!ht]
\includegraphics[width=0.6\linewidth] {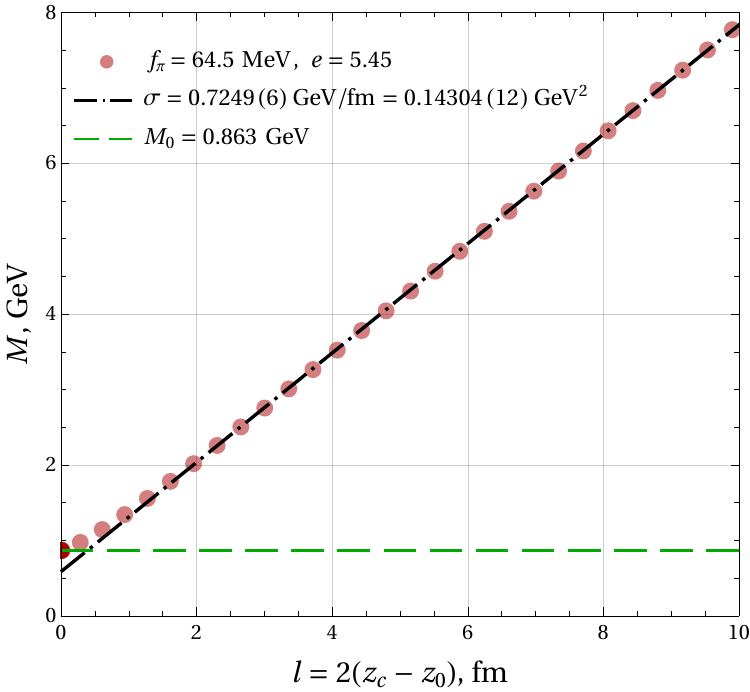}
  \caption{Dependence of the static mass $M=-L$~(\ref{lagrF}) on the longitudinal deformation parameter $l = 2 (z_c - z_0)$, where $z_c$ is the center of inertia of half of the object~(\ref{eq:center_of_inertia}) and $z_0$ - the center of inertia at $a \to 0$. $M_0$ is the mass of the static solution at $a \to 0$.}
  \label{fig_M_on_l}
\end{figure}

Following \cite{Adkins_1983} and choosing $f_\pi = 64.5\;\mathrm{MeV}$ and $e = 5.45$,  we can then obtain the tension coefficient of the effective string $\sigma \approx 0.14~\mathrm{GeV}^2$, which is consistent with various models of confinement~\cite{quigg_1979,  eichten_1980, kogut_1991, bali_2001}. The dependence of the mass on the longitudinal deformation parameter~$l$ (in fermi units) is shown in Fig.~\ref{fig_M_on_l}. As $a \to 0$, we obtain a spherically symmetric case with the value of the static mass $M_0 = 863$~MeV.

\section{Conclusions}
The analysis of the modified Skyrme model with the Cassini ansatz convincingly confirms the effectiveness of this approach for describing confinement mechanisms in quantum chromodynamics. 

Deformation of a spherically symmetric object leads to the creation of a skyrmion–anti-skyrmion pair in the internal region of Cassini ovals with a vanishing baryon number in sum. Conversely, the initial baryon charge remains in the external region. The total charge remains conserved throughout the deformation and determined by the outer region. Remarkably, the baryon charge does not depend on the explicit form of $F(\xi)$, only the boundary values matter.

The main achievement of the work was obtaining a linear dependence of static energy on the longitudinal deformation parameter with slope $\sigma \approx 0.14\ \mathrm{ GeV}^2$, which qualitatively corresponds to the string tension coefficient, both in various phenomenological models of confinement and in lattice quantum chromodynamics. 
The result demonstrates the key role of the topological properties of the model in forming the Cornell potential.

\vspace{2ex}
\begin{acknowledgments}
The authors thank Professor Vladimir Andreevich Nikolaev for fruitful discussions of the results obtained. NG and AM are supported by the state assignment of the Ministry of Science and Higher Education of Russia (Project No. FZNS-2024-0002). Eternal thanks to Oleg Grigorievich Tkachev, without whom this project would not have been possible. 
\end{acknowledgments}

\bibliography{cassini}

\end{document}